\documentclass[journal]{IEEEtran}
\usepackage{graphicx}
\usepackage{subfigure}

\begin{document}
\title{Development of CMOS monolithic pixel sensors with
in-pixel correlated double sampling\\ and fast readout}
%
%

\author{Marco Battaglia,
        Jean-Marie Bussat,
        Devis Contarato,
        Peter Denes,
        Piero Giubilato,
        Lindsay E. Glesener
\thanks{Manuscript received May 30, 2008.
We thanks the staff of the LBNL ALS and 88'' cyclotron and of the FNAL MTBF for 
their assistance during data taking and the excellent performance of the accelerators.
This work was supported by the Director, Office of Science, of the
U.S. Department of Energy under Contract No.DE-AC02-05CH11231.}
\thanks{M. Battaglia is with the Department of Physics, University of 
California and the Lawrence Berkeley National Laboratory, Berkeley, CA 94720, 
USA (telephone: 510-486-7029, e-mail: mbattaglia@lbl.gov).}%
\thanks{J.M. Bussat was with the Lawrence Berkeley National Laboratory, 
Berkeley, CA 94720, USA, now is with Stanford University, CA 94305, USA}%
\thanks{D. Contarato is with the Lawrence Berkeley National Laboratory, 
Berkeley, CA 94720, USA}%
\thanks{P. Denes is with the Lawrence Berkeley National Laboratory, 
Berkeley, CA 94720, USA}%
\thanks{P. Giubilato is with the Istituto Nazionale Fisica Nucleare, Sezione 
di Padova, Italy and a visitor at the Lawrence Berkeley National Laboratory, 
Berkeley, CA 94720, USA}%
\thanks{L.E. Glesener is with the Department of Physics, University of 
California and the Lawrence Berkeley National Laboratory, Berkeley, CA 94720, USA}%
}

\maketitle
\pagestyle{empty}
\thispagestyle{empty}

\begin{abstract}
This paper presents the design and results of detailed
tests of a CMOS active pixel chip for charged particle
detection with in-pixel charge storage for correlated
double sampling and readout in rolling shutter mode at
frequencies up to 25~MHz. This detector is developed in 
the framework of R\&D for the Vertex Tracker for a future
$e^+e^-$ Linear Collider.

\end{abstract}


\section{Introduction}

\IEEEPARstart{T}{he} Vertex Tracker for a future $e^+e^-$ Linear Collider, such as the 
International Linear Collider (ILC) project, 
has requirements in terms of position resolution and material budget 
that largely surpass those of the detectors at LEP, SLC and LHC. 
The single point resolution needs to be $\le$~3~$\mu$m and the detector has be read-out 
fast enough that the machine-induced background does not adversely affect the track 
pattern recognition and reconstruction accuracy. Detailed simulation of incoherent pair 
production in the strong field of the colliding beams, the dominant background source, 
predicts a hit density of order of 5~hits~bunch~crossing$^{-1}$~cm$^{-2}$ on the innermost 
detector layer, located at a radius of 1.5~cm from the interaction point, for 250~GeV beams 
and a solenoidal field of 4~T. 
The requirement of a hit occupancy $\le$0.1~\%, needed to ensure clean standalone pattern 
recognition in the Vertex Tracker, corresponds to a maximum of 80 bunch crossings 
which can be integrated  by the detector in a readout cycle, i.e.\ a readout frequency of 
25~MHz for a detector with 512 pixel long columns.
Finally, power dissipation must be kept small enough, so that cooling can be achieved 
by airflow, without requiring active cooling systems that contribute significantly to 
the material budget of the vertex detectors installed in the LHC experiments. 
Tests performed on a carbon composite prototype ladder, equipped with 50~$\mu$m-thin CMOS 
pixel sensors, have shown that an airflow of 2~m~s$^{-1}$ can remove 
$\simeq$ 80~mW~cm$^{-2}$. Assuming a pixel column made of 512 pixels, this 
corresponds to a maximum allowable power dissipation of $\simeq$0.5~mW~column$^{-1}$.

An attractive sensor architecture for the Linear Collider Vertex Tracker sensor 
is a pixel of $\simeq$20$\times$20~$\mu$m$^2$ readout at 25-50~MHz during 
the long ILC bunch train. Signals are digitised at the end of the column with 
enough accuracy to allow charge interpolation to optimise the spatial resolution.
Digitising at the required speed and within the maximum tolerable power dissipation 
poses major design challenge. The study of data collected with a CMOS pixel test 
chip, with 10$\times$10~$\mu$m$^2$, 20$\times$20~$\mu$m$^2$ and 
40$\times$40~$\mu$m$^2$ pixels~\cite{snic}, shows that a 5-bit ADC accuracy is 
sufficient, provided that pixel pedestal levels are subtracted before digitisation. 
This requires performing pedestal subtraction either in pixel or at the end of the 
column.

\section{LDRD-2: A pixel chip with in-pixel CDS}

We have designed a CMOS monolithic pixel chip with in-pixel correlated 
double sampling (CDS) and tested it for readout speeds up to 25~MHz. 
The chip consists of a matrix of 96$\times$96 pixels arrayed on a 20~$\mu$m 
pitch. Each pixel has two 5$\times$5~$\mu$m$^2$ polysilicon-insulator-polysilicon 
(PIP) capacitors, corresponding 
to a capacitance of $\simeq$20~fF. These capacitors are used for the storage of 
the pixel reset and signal levels. The net pixel signal is obtained by 
subtracting the reset from the signal level. In the current CDS implementation, 
this subtraction is performed off-chip, allowing for a detailed performance study.
Pixels are readout in rolling shutter mode, which ensures a constant integration 
time across the pixel matrix. The pixel array is divided in two 48$\times$96 pixel 
sections, which are readout in parallel. Different pixel designs, including 
diode sizes of 3~$\mu$m and 5~$\mu$m, have been implemented. One half of the pixel 
matrix implements charge-collection diodes surrounded by a guard-ring in order to 
study the effect of a guard-ring after irradiation.

The detector has 
been fabricated in an AMS 0.35~$\mu$m 4-metal, 2-poly CMOS-OPTO process, which 
provides an epitaxial layer with a nominal thickness of 14~$\mu$m.

The readout sequence is as follows. On a reset signal on the pixel $i$, 
the pixel reset level is stored and the charge is integrated on the diode, while the 
reset level is stored for the pixel $i+1$. After the integration time which corresponds
to the scan of a section, i.e.\ $N_p/f$, where $N_p$ is the number of pixels in a readout 
section and $f$ the readout frequency, the signal level is stored, the pixel is reset and 
the cycle restarted. The reset and signal levels are readout serially. The highest 
frequency tested is 25~MHz, corresponding to an integration time of 184~$\mu$s. 
In order to minimise the power dissipation of the pixel cell, the source follower 
is switched on only in the short time elapsed between the write signal level of one 
event and the write reset level of the next event.

The detector is readout through a custom FPGA-driven acquisition board. The board 
has four 14~bit, 40~MSample/s TI ADS5421 ADCs which read the chip outputs, while an 
array of digital buffers drives all the required clocks and synchronisation signals. 
For each of the 48$\times$96 pixel halves, the signal and reset levels stored in the 
in-pixel capacitors are sent to two different analog outputs which are subsequently 
digitised by two different ADCs on the readout board. 
The FPGA has been programmed to generate the clock pattern and collect the sampled 
data from the ADCs. A 32~bit wide bus connects the FPGA to a NI-6533 digital acquisition 
board installed on the PCI bus of the control PC. Data are processed on-line
by a LabView-based program, which performs the subtraction of the reset level 
from the pixel level and computes the pixel noise and residual pedestal.  

\section{Pixel Response Tests}

The detector has been tested in the lab using a 2.2~mCi $^{55}$Fe collimated 
source. The detector performance has been studied as a function of the readout 
frequency, from 1.25~MHz up to 25~MHz. First the pixel noise has been measured 
for operation at room temperature. No significant degradation of the noise of 
the pixel matrix is observed up to the highest frequency. The measured noise is due, 
in part, to the readout electronics noise, which has been estimated to be 
$\simeq$20~ENC. The pixel noise has been studied as a function of the operating 
temperature (from +20~$^o$C to -20~$^o$~C). We measure (42$\pm$4)~ENC, (54$\pm$7)~ENC 
and (54$\pm$6)~ENC at +20~$^o$C and (37$\pm$6)~ENC, (43$\pm$5)~ENC and (46$\pm$5)~ENC 
at -20~$^o$C, at readout frequencies of 1.25~MHz, 6.25~MHz and 25~MHz, respectively,
where the quoted uncertainties represent the spread of the noise measured on the 
pixels of a sector.

\begin{figure}[!ht]
\centering
\includegraphics[width=3.0in]{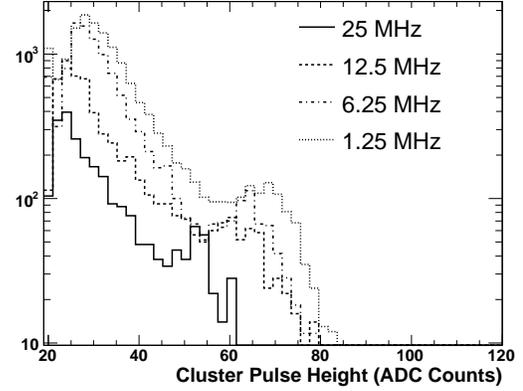}
\caption{Response of the LDRD-2 chip to 5.9~keV X-rays from $^{55}$Fe for readout 
frequencies of 1.25~MHz (dotted),  6.25~MHz (dash dotted), 12.5~MHz (dashed) 
and 25~MHz (continuous).}
\label{fig:55fe}
\end{figure}  

The pixel calibration has been obtained using the 5.9~keV X-rays from a $^{55}$Fe 
source for various readout frequencies. Charge generated by X-rays which convert in 
the shallow depletion region near the pixel diode is fully collected, resulting in a
pulse height peak corresponding to the full X-ray energy, which generates on average 
1640 electrons in the Si sensitive volume. We select 
pixel clusters consisting of either a single pixel with a pulse height in excess of 
seven times the pixel noise or a pixel fulfilling the same requirement plus an 
adjacent pixel with pulse height in excess of three times the pixel noise. 
The resulting pulse height spectra are shown in Figure~\ref{fig:55fe}, for different 
readout frequencies.
\begin{figure}[!hb]
\centering
\includegraphics[width=3.5in]{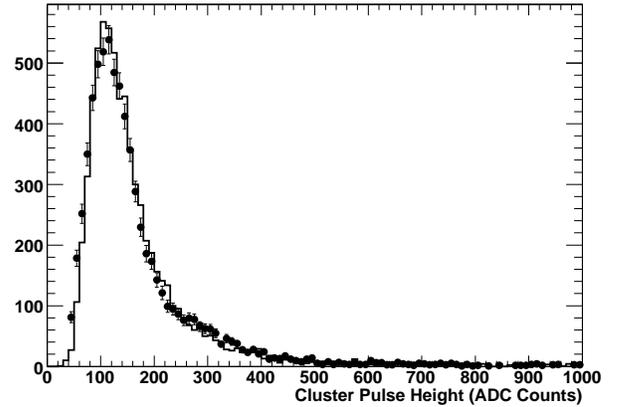}

\includegraphics[width=3.5in]{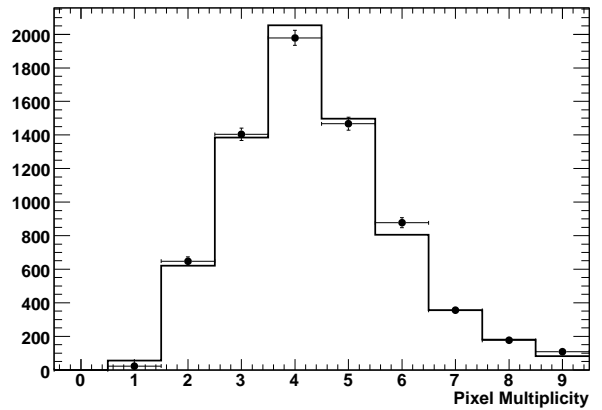}
\caption{Response of the LDRD-2 pixel chip to 1.35~GeV $e^-$: Data (points with error
bars) are compared to predictions from our simulation based on {\sc Geant-4} and a 
dedicated sensor simulation in {\tt Marlin} (line) for the cluster pulse height 
(above) and pixel multiplicity in the cluster (below).}
\label{fig:sim}
\end{figure}  

The response to high momentum particles has been studied in beam tests. 
We used both the 1.35~GeV electron beam from the LBNL Advanced Light 
Source (ALS) booster and the 120~GeV secondary proton beam at the Meson Test 
Beam Facility (MTBF) at Fermilab, as part of the T966 beam test 
experiment~\cite{Battaglia:2008nj}. 
\begin{figure}[!t]
\centering
\includegraphics[width=3.75in]{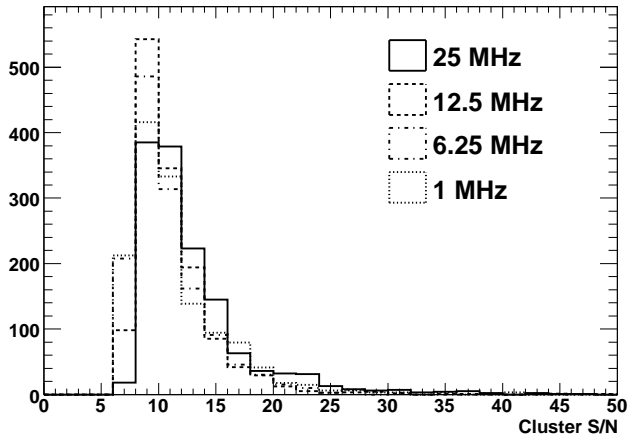}
\caption{Response of the LDRD-2 chip for various 
readout frequencies: cluster signal-to-noise distributions 
for 1.35~GeV $e^-$s for readout 1~MHz (dotted), 6.25~MHz (dash dotted), 
12.5~MHz (dashed) and 25~MHz (continuous).}
\label{fig:sn}
\end{figure}  
Data are converted into the {\tt lcio} format~\cite{Gaede:2003ip}, which is the 
persistency format adopted by the ILC studies. The data analysis is performed 
offline by a dedicated set of processors developed in the {\tt Marlin} C++ 
framework~\cite{Gaede:2006pj}.
Events are first scanned for noisy pixels. The noise and pedestal values 
computed on-line are updated, using the algorithm in ~\cite{chabaud}, in order 
to follow possible variations in the course of a data taking run.

A cluster search is performed next. Each event is scanned for pixels with pulse 
height values over a signal-to-noise (S/N) threshold of 5, these are designated 
as cluster `seeds'.  Seeds are then sorted according to their pulse heights and 
the surrounding, neighbouring pixels are tested for addition to the cluster. 
\begin{figure}[!h]
\centering
\includegraphics[width=3.25in]{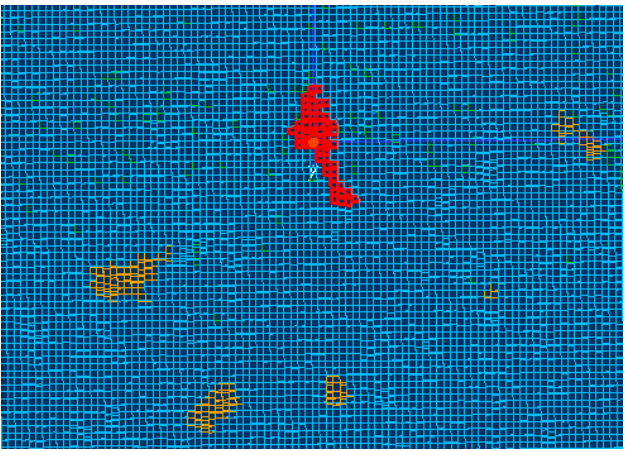}
\caption{LDRD-2 response to low momentum electrons. This event display shows 
clusters produced by low momentum electrons produced by the interaction of 1.35~GeV 
electrons on an Al scraper. Clusters are evidently broad and asymmetric.}
\label{fig:lowpe}
\end{figure}
The neighbour search is performed on a 7$\times$7 matrix surrounding the pixel 
seed and the neighbour threshold is set at 2.5, in units of the pixel noise.     
Clusters are not allowed to overlap, i.e. pixels already associated to one 
cluster are not considered for populating another cluster around a different 
seed. Finally, we require that clusters are not discontinuous, i.e. pixels 
associated to a cluster cannot be separated by any pixel below the 
neighbour threshold. The expected pixel response is simulated with a dedicated 
digitisation processor in {\tt Marlin}~\cite{pixelsim}. The charge collection process 
is described starting from ionisation points generated along the particle trajectory 
using {\tt Geant~4}~\cite{Agostinelli:2002hh}, by modelling the diffusion of charge 
carriers, originating in the epitaxial layer, to the collection diode. Simulated 
data are then processed through the cluster reconstruction stage, using the same 
processor as beam test data. A comparison of the real and simulated cluster pulse 
height and pixel multiplicity obtained, after tuning the charge carrier diffusion 
length in simulation to provide the best fit to the pixel multiplicity 
distribution, is shown in Figure~\ref{fig:sim}
\begin{figure*}[!th]
\centerline{\includegraphics[width=5.75in]{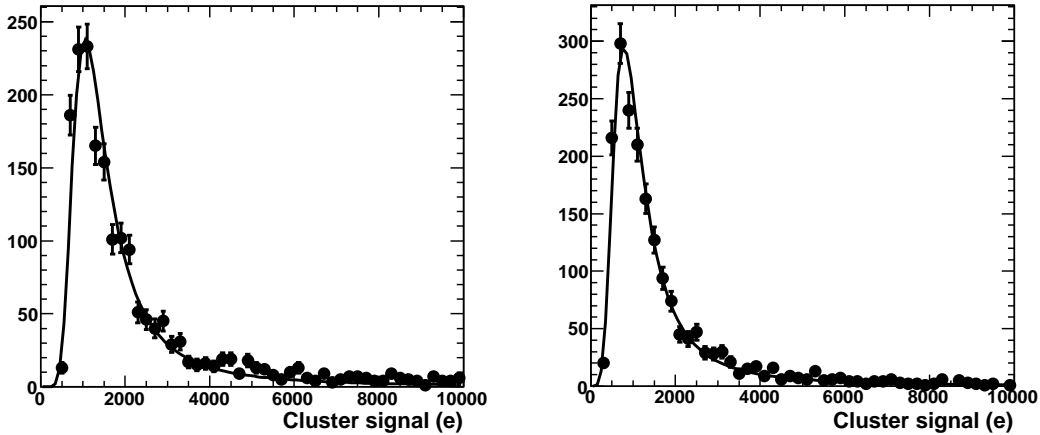}}%
\caption{Response of the LDRD-2 pixel chip to 120~GeV protons: 
Cluster pulse height distributions obtained for pixels with 5~$\mu$m (left) 
and 3~$\mu$m (right) diode. The continuous lines show the 
interpolated Landau functions.}
\label{fig:landau}
\end{figure*}
Finally, the response to low momentum electrons has been studied. The response
of the detector to low energy electrons is important since particles from pair 
background have energies typically below 100~MeV. Data has been collected at the 
ALS with an Al beam scraper placed a few meters upstream of the LDRD-2 detector. 
A large fraction of the registered hits are due to low energy electrons, originating 
from the interaction of the primary beam with the Al scraper, or to tertiaries,
from photon conversions. These hits are characterised by large, asymmetric clusters
(see Figure~\ref{fig:lowpe}) with average multiplicity exceeding 1.5 times that of 
clusters from high energy electrons. This suggest that low-energy background hits 
could be identified and rejected, based on the observed cluster shape.

The response to high energy hadrons has been studied with 120~GeV protons at MTBF.
The most probable cluster pulse height has been measured to 
be (1112$\pm$17)~$e^-$s for 5~$\mu$m and (830$\pm$13)~$e^-$s for 3~$\mu$m collecting 
diodes as shown in Figure~\ref{fig:landau}. The most probable pixel multiplicity 
scales from 2.7 to 4.3, respectively. An average signal-to-noise ratio of 12 to 
13 has been measured for 1~MHz and 25~MHz readout, respectively 
(see Figure~\ref{fig:sn}). The detector was operated at +20$^{\circ}$C and the 
signal-to-noise performance is limited in part by the noise of the read-out board, 
as discussed above.

\section{Radiation Hardness Tests}

At an $e^+e^-$ linear collider there are two main sources of radiation, which need 
to be considered to assess the radiation tolerance of the sensor. The first is due 
to low energy neutrons produced by giant resonance excitations in the beam dump 
and in the interaction of low energy pair background on the forward masks. 
The neutrons at the location of the Vertex Tracker has been estimated to 
generate a fluence $\simeq$6~$\times$~10$^{10}$~n~cm$^{-2}$~year~$^{-1}$. The 
second radiation source is due to low momentum pairs produced in the electromagnetic 
interaction of the colliding beams corresponding to a dose of $\simeq$~50~kRad/year
on the Vertex Tracker.

\begin{figure}[!h]
\centering
\includegraphics[width=3.75in]{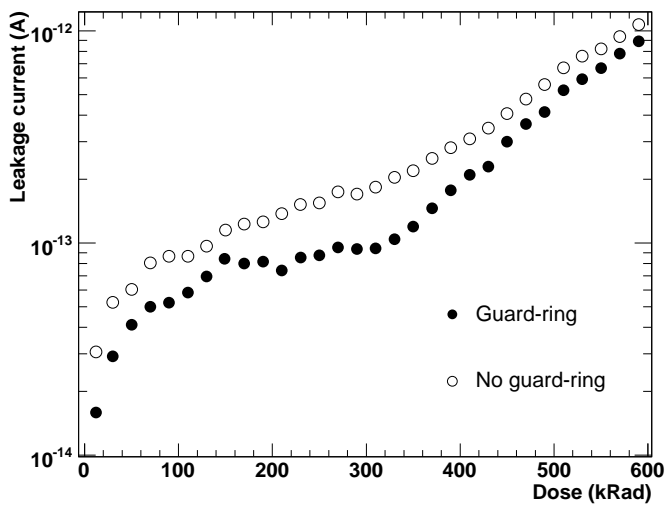}
\caption{Results of the LDRD-2 sensor irradiation with 200~keV electrons. Sensor 
leakage current as a function dose for pixel cells designed with (filled dots) 
and without (open dots) guard ring around the charge-collecting diode.}
\label{fig:radel}
\end{figure}

The effect of ionising radiation has been assessed  by exposing the chip to 
200~keV electrons at the 200~CX electron microscope of the National Center 
for Electron Microscopy at LBNL. Electrons of 200~keV are below the displacement
damage threshold for Si.  The sensor has been irradiated with a flux of 
$\simeq$~2300 $e^-$ s$^{-1}$ $\mu$m$^{-2}$, in multiple steps. In between 
consecutive irradiation steps, 100 events are acquired without beam and the pixel 
pedestals and noise computed, in order to monitor the evolution of the pixel leakage 
current with dose. Figure~\ref{fig:radel} shows the pixel pedestal levels, which 
measure the leakage current, as a function of the integrated dose. All tests have been 
performed at room temperature. 
Results show that the pixel functions properly up to several hundreds kRads, i.e.\
doses comparable to five or more year of operation at a linear collider, with 
leakage currents below 1~pA which is an acceptable level and can be further 
controlled by cooling the device during operation. 
The effect of a guard-ring around the charge-collecting diode is also apparent.

The effect of non-ionising radiation has been studied exposing the 
LDRD-2 detector chip to neutrons produced from 20~MeV deuteron breakup 
on a thin target at the LBNL 88-inch cyclotron~\cite{mcmahan}. 
The detector chip was located 8~cm downstream from the target and 
activation foils were placed just behind it to monitor the fluence. Deuteron 
breakup produces neutrons on a continuum spectrum from $<$~1~MeV up to 
$\sim$~14~MeV. A beam current of 800~nA was used, corresponding to an estimated 
flux at the chip position of $\simeq$~4~$\times$ 10$^{8}$~$n$~cm$^{-2}$~s$^{-1}$.  
We compared the pixel noise before and after irradiating the chip with a total 
fluence of 1.2$\times$10$^{13}$~$n$~cm$^{-2}$. As the 
observed increase in noise of (7$\pm$8)~$e^-$ is not significant, the 
test shows that the chip can withstand neutrons fluxes significantly above 
those foreseen at the interaction region of an $e^+e^-$ linear collider.

\section{Conclusions and Outlook}

The LDRD-2 pixel cell has small pitch and in-pixel charge storage for correlated 
double sampling. This offers a viable architecture for a pixel sensor with analog
readout able to match the timing requirements of the Vertex Tracker at a linear 
collider such as the ILC.
The chips has been characterised using $^{55}$Fe and particle beams.
The chip performance is found to be stable up to the highest tested readout frequency 
of 25~MHz, corresponding to an integration time of 184~$\mu$s. The radiation tolerance 
of the pixel cell has been tested with low-energy electrons and neutrons. The pixel 
withstands radiation fluxes in excess to those expected after several years of operation 
in the Vertex Tracker of an $e^+e^-$ linear collider.
The pixel cell is the basis of a third-generation chip, which further addresses the 
requirements of the ILC, or other future $e^+e^-$ linear colliders. 
The LDRD-3 chip consists of a matrix of 96$\times$96 pixels on a 20~$\mu$m pitch, 
with the same pixel cell as the LDRD-2. In addition, the chip features a column parallel 
readout at frequency up to 50~MHz and the digitisation is performed on-chip, at the end 
of each column, by a row of successive approximation, fully differential ADCs featuring 
low power dissipation. Each ADC has a size of 20~$\mu$m$\times$1~mm, which matches the 
pixel pitch. The digitisation of the 96 rows is performed in 1.9~$\mu$s.

\end{document}